\documentclass{amsart}
\usepackage{graphicx,amsmath}
\usepackage{color}

\newcommand{\rv}{r} 

\newcount\n  \newdimen\w

\def\Repeat#1#2{\n=#1\relax\loop\ifnum       
  \n>0\relax #2\advance\n by-1\repeat}

\long\def\OMIT#1{\relax }  

\def\re#1{(\ref{#1})}   
  
\def\eqn#1#2{ \begin{align} \label{#1}         #2 \end{align}}

\def\nl#1{          \\ \label{#1}        }  

\def\delim#1#2#3{\csname\ifcase#1 relax\or   
   big\or Big\or bigg\or Bigg\fi\endcsname   
  {\ifcase#2\or\Delim#3\or\deliM#3\fi}}      
\def\Delim#1{\ifcase#1\relax\or(\or[\or\{\or<\or\langle\or|\or\|\or---{ }\fi}
\def\deliM#1{\ifcase#1\relax\or)\or]\or\}\or>\or\rangle\or|\or\|\or{ }---\fi}

\let\f\frac                     
        
\begin{document}

\title{Non-equilibrium thermodynamical framework for rate- and state-dependent 
friction}
\author{P. V\'an$^{1,2}$, N. Mitsui$^{1}$ and T. Hatano$^3$}
\address{$^1$Dept. of Theoretical Physics, Wigner RCP, RMKI, \\  H-1525 Budapest, P.O.Box 49, Hungary; 
 {$^2$Dept. of Energy Engineering, Budapest Univ. of Technology and 
 Economics},\\
  H-1111, Budapest, Bertalan Lajos u. 4-6,  Hungary,\\
  {$^3$Earthquake Research  Institute, University of Tokyo, Japan}}

\date{\today}
\markleft{Thermodynamics of friction}

\begin{abstract}
Rate- and state-dependent friction law for velocity-step and healing are 
analysed from a thermodynamic point of view. Assuming a logarithmic deviation 
from steady-state a unification of the classical Dieterich and Ruina models of 
rock friction is proposed. 
\end{abstract}

\maketitle

\section[intro]{Introduction}

The rock experiments of sliding friction are understood by the so-called 
rate- and state-dependent friction laws. The equations of this laws unify the 
results 
obtained from two types of rock experiments; the first one is the time 
dependence of static coefficient of friction \cite{Die72a} and the second 
one is slip velocity dependence of the dynamic coefficient of friction 
\cite{Die78a}.

The properties of dynamic friction are the following \cite{Maro98a}: 
\begin{enumerate}
 \item frictional coefficient in stable sliding conditions with a constant load-point velocity depends on the logarithm of the load-point velocity;
 \item the magnitude of the instantaneous jump of the frictional coefficient 
 depends on the change of the logarithm of the quotient of the corresponding 
 load-point velocities;
 \item the following evolution of the frictional coefficient to new value in 
 stable sliding is also dependent on the instantaneous change of the load-point 
 velocity; 
\item oscillation occurs in some cases (e.g., large load-point velocity, 
polished surfaces, thin sand interface layer between the samples) (see e.g., 
\cite{MarEta90a}).
\end{enumerate}

In healing experiments the properties of static friction are: 
\begin{enumerate}
 \item recovery magnitude is proportional to the logarithm of healing time;
 \item larger velocity or larger elasticity increases the recovery magnitude of 
 the static friction;
\end{enumerate}
  
  
These properties can be reproduced by using two classical equations. The first 
one is the {\em constitutive law} \re{Die_flaw}, expressing the relation 
between frictional coefficient $\mu$ and slip velocity $V$ with an additional 
variable, called state variable $\theta$. The second one is the {\em evolution 
law} \re{Die_elaw} expressing the time evolution of state variable depending on 
the slip velocity \cite{Die79a}:
\begin{eqnarray}
 \mu = \frac{\tau}{\sigma} &=& 
    \mu_* + a\ln\left( \frac{V}{V_0} \right)\ + b\ln\left( 
    \frac{V_{0}\theta}{L} \right), \label{Die_flaw}\\
 \frac{d\theta}{dt} &=& 
    1-\frac{V\theta}{L},
\label{Die_elaw}\end{eqnarray}
where $\tau$ is the shear stress, $\sigma$ is the normal stress, $\mu_{*}$ is 
the  constant frictional coefficient for steady-state slip at reference slip 
velocity $V_{0}$, $a$ and $b$ are material parameters, $L$ is the 
critical slip distance, and $t$ is time. 

An important improvement is the evolution equation of Ruina \cite{Rui83a}.
\eqn{Rui_elaw}{
 \frac{d\theta}{dt} = 
    -\frac{V\theta}{L} \ln\left(\frac{V\theta}{L} \right).
}

Experimental data of static friction is better reproduced by the equations of 
the Dieterich-law \cite{Die79a} (eqs. \re{Die_flaw}  and \re{Die_elaw}), and  
of dynamic friction by the equations of  Ruina-law \cite{Rui83a} (eqs. 
\re{Die_flaw} and \re{Rui_elaw}). A comparison with experimental data is given 
in the works \cite{Maro98a,NagEta12a}. In  particular the Dieterich-law 
assumes time-relaxation and therefore it is  asymmetric for upward and downward 
jumps in displacement, contrary to the  experiments. On the other hand the 
experimentally observed time dependent healing is properly reproduced by the 
Dieterich-law and does not reproduced by the Ruina-law.  Thus another versions 
have been proposed (e.g., \cite{PerEta95a,KatTul01a,NagEta12a}) in order to 
reproduce the experimental data better. However, none of them are completely 
satisfactory.

Nakatani reformulated the Dieterich-Ruina law introducing a new variable 
\cite{Nak01a}:
\eqn{Nak_var}{
\theta = \f{L}{V_0} \exp\left(\f{\Theta}{b}\right).
}

Then the constitutive law is linearized with this variable:
\eqn{Nak_claw}{
 \mu = \frac{\tau}{\sigma} = 
    \mu_* + a\ln\left( \frac{V}{V_0} \right)+\Theta,
}
and the evolution laws of Dieterich and Ruina become
\eqn{Nak_DRelav}{
\frac{d\Theta}{dt} &= \frac{b}{L}\left(V_0 e^{-\f{\Theta}{b}}-V\right),
\nl{Nak_Relaw}
  \frac{d\Theta}{dt} &=  - V\left(\f{b}{L} \ln\left(\f{V}{V_0}\right)+
  \f{\Theta}{L}\right),
}
respectively. Nakatani suggested this modification together with a particular 
interpretation of the new variable as strength and  interpreted the modified 
evolution equation of Dieterich in the framework of thermal activation 
theory. He did not investigate the modified form of Ruina law \re{Nak_Relaw}.

\section{Thermodynamics of the frictional layer}

A thermodynamic approach of Mitsui and V\'an introduced a minimal model of rock 
friction by using permanent and recoverable parts of the total 
observed displacement as state variables in the spirit of continuum plasticity 
\cite{MitVan14a}. 
They interpreted the state variable of the rate- and state dependent friction 
law as the elastic part of the displacement. It was argued that the common 
origin of the constitutive and the evolution laws is originated in a 
thermodynamic framework. Their calculation of the entropy production was the 
following.

The general setting is a sliding body on a horizontal surface with mass $m$. 
There are two 
forces that determine the motion of the body: the external force $F_e$, and the 
damping force $F_d$, due to friction (Fig. \re{fig_slidingbody}). The position 
of the body is denoted by $x$. The body is not considered completely rigid, 
however one assumes that one particular material point of the body  
characterize its instantaneous position. The equation of motion is
\eqn{Newt}{
  m\ddot x = F_e - F_d.
}
\begin{figure}
\centering
\includegraphics[width=11cm,height=8cm]{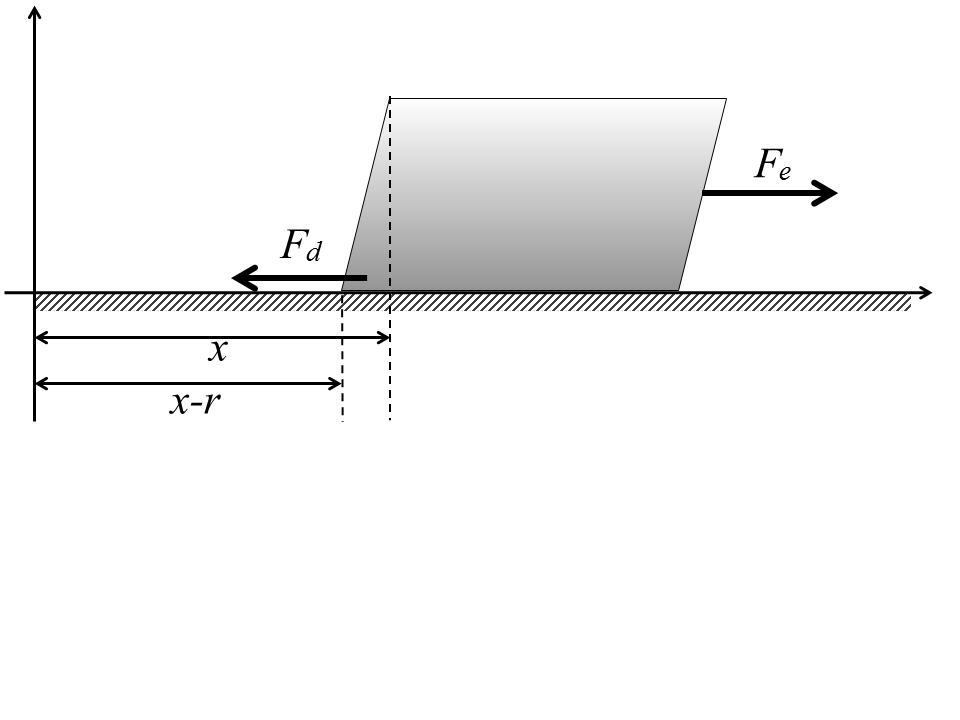}
\caption{ \label{fig_slidingbody}
 Sliding thermomechanical body
}
\end{figure}
Moreover, the work of the external force changes the energy of the body, $E$. 
Therefore  
\eqn{Etot}{
  \dot E = F_e \dot x.
}

In this case thermodynamics requires that the damping force contributes only to 
the internal energy of the body. 
It is assumed that the external force accelerates the body 
and also that the body is deformable. In this homogeneous model, the 
deformation is expressed by the recoverable displacement, $r$. Therefore the 
kinetic and and elastic energies of the body are distinguished. This 
particular interpretation from \cite{MitVan14a} is not necessary, $\rv$ may 
denote a general internal variable.

The internal energy, $U$, is the difference of the total energy, $E$, the 
kinetic energy and a quadratic contribution of the internal variable.  This 
form follows from the condition of thermodynamic stability in the state space 
\cite{Ver97b}.
\eqn{Eint}{
U=E-m\frac{{\dot x}^2}{2}-k\frac{\rv^2}{2},
}
where  $k$ is a parameter like in elasticity.  
One assumes a particular kinematic condition in order to introduce an 
interpretation of $\rv$. When the  instantaneous position of the body is the 
sum of a permanent and a recoverable displacements then a convenient method of 
their distinction is an additive separation of the displacement rates:
\eqn{plastkin}{
  V= \dot x = \dot \rv + z,
}
where $V$ is the rate of the position $x$,  and $z$ is the rate of the 
permanent displacement. This rate type kinematical condition is convenient when 
distinguishing permanent and recoverable changes. \re{plastkin} is analogous to 
the condition used in plasticity for the distinction of plastic and elastic 
strains (see e.g., \cite{FulVan12a,RusRus11b}). However, in  friction the 
internal variable is not necessarily identical to the recoverable strain. 

The dissipation can be calculated by the entropy balance, assuming that the entropy is the function of the internal energy only:
\eqn{Sbal}{
\dot S(U)=\frac{1}{T}\left( F_e \dot x -mV\dot V-k\rv\dot \rv  \right)\geq0\,\ 
\Rightarrow\,T\dot S =F_d V-kr\dot r\geq0.
}

The damping force and also the rate of the internal variable $\dot r$ 
are the constitutive quantities to be determined in accordance with the 
requirement of nonnegative entropy production. We connect the first term to 
friction interaction and the second term to frictional healing.
In the vicinity of thermodynamic equilibrium, where both forces and fluxes 
are zero, the usual linear relationship is a consequence of Lagrange's mean 
value theorem \cite{Gur96a,Van03a}.  $V$ and $r$ are thermodynamic state 
variables, therefore they can be considered as thermodynamic forces, while 
$\mu$ and $\kappa\dot r$ are to be determined constitutively, they are 
thermodynamic fluxes. The standard linear approximation results in 
equations that reflect well the thermodynamic admissibility of both 
velocity weakening and strengthening, however, did not incorporate the observed 
direct effect (logarithmic relaxation) \cite{MitVan14a}.

However, in velocity-step experiments dynamic friction is a 
steady state phenomenon, we are looking for a 
deviation of the frictional coefficient from a fixed value, $\mu_0$, at a given 
$V_0$ reference velocity. On the other hand the internal variable is 
expected to be zero, when its time derivative is zero, in  
thermodynamic equilibrium. Therefore we face to a mixed, partial 
steady state, partial equilibrium situation. However, the dissipation 
inequality \re{Sbal} may not characterize the deviation from steady state. What 
we need is an estimation of the deviation of the entropy production. 

In our framework the steady state means a constant displacement rate, denoted 
by $V_0$. Therefore the internal energy should be determined accordingly 
\eqn{Eintm}{
U=E_0-m\frac{(V-V_0)^2}{2}-k\frac{\rv^2}{2},
}
where $E_0$ is the energy of the body moving with the veolcity 
$V_0$\footnote{The proper relation of kinetic and internal 
energies in a Galilei relativistic framework is a delicate question from a 
thermodynamic point of view. A detailed treatment of objective, frame 
indpendent thermodynamic modeling in case of single component fluids is given 
in \cite{Van15a}. A similar approach is required in our case, too}.
Then the calculation of the entropy production results in 
\eqn{Sbalm}{
T\dot S(U)= F_e V -m(V-V_0)\dot V-k\rv\dot \rv 
 =F_e V_0 + F_d(V-V_0)-kr\dot r\geq0.
}

We can introduce a friction specific entropy production with Amonton's law:
\eqn{fSbal}{
\f{T}{F_n A}\dot S= \Sigma =  \mu_0 V_0+\Delta \mu (V-V_0)-\kappa r\dot r\geq0,
}
where $\mu_0 = F_e/(A F_n)$ is the external shear stress divided by the normal 
force, $\Delta\mu = F_d/(A F_n)$ is the frictional stress divided by the 
normal force and $\kappa = \f{k}{A F_n}$. 

Now, it is straightforward to introduce a linear approximation for the 
increment of entropy production, as for the near equilibrium situation. 
However, in the following we apply a different starting point. Our fundamental 
assumption is that the leading term of the deviation  from the steady state is 
{\em logarithmic}, while it is linear around equilibrium. We can formulate this 
hypothesis analogously to the classical exploitation of the entropy principle 
introducing  the concept of {\em incremental entropy production}. We require 
that it is minimal at the steady state 
\eqn{ss_epr}{
\Delta \Sigma = \Sigma - \Sigma_{steady} = \Sigma- \mu_0 V_0 = 
\Delta \mu \ln\left(\f{V}{V_0}\right) - \kappa r\dot r \geq 0
}
We will call this hypothesis {\em the principle of minimal incremental entropy 
production}. 

The required minimality ensures the asymptotic stability of the steady state if 
the evolution equations are constructed accordingly. In this respect the 
principle is similar to the role 
of the second law near to the equilibrium. On the other hand this hypothesis is 
a modification of the requirement of nonnegative excess entropy production of 
Glansdorff and Prigogine \cite{GlaPri71b}. The difference of the 
Prigogine-Glansdorff requirement is the logarithmic deviation 
instead of a linear one. (The possibility of partial steady state, a specific 
property of friction, is important, too). Along with them we want to emphasize 
that the inequality here is not a law of nature, it is regarded as a 
convenient stability assumption \cite{KeiFox74a,GlaEta74a}.

An example of similar logarithmic deviation is the thermodynamics of 
chemical kinetics, where the entropy production is a product of the chemical 
affinity and the reaction rate, but the Guldberg-Waage kinetic equations 
introduce an exponential relation \cite{GroMaz62b}. It is 
remarkable that chemical equilibrium is considered as a steady state from a 
thermodynamic point of view, when forward 
and backward reactions are properly distinguished \cite{LenGya81a,Len89a}. 
Our formula corresponds a simple monomolecular reaction, which is 
actually equivalent to a single internal degree of freedom \cite{GroMaz62b}. 
In friction the logarithmic deviation may be further motivated 
according to this chemical analogy \cite{Nak01a}. 

Finally we remark, that the logarithmic form can be directly derived assuming 
that the internal energy is modified by a logarithmic velocity dependent term 
instead of a quadratic one $U= E_0 - m V(Log(V/V_0)-1)$.

\re{ss_epr} is similar to the usual entropy production in many respect. First 
of all the two terms are zero in the reference steady state and we assume 
that $\Delta\mu$ is a constitutive function of the logarithmic  
deviation, a function of the thermodynamic state variable. Therefore, in case 
of smooth functions the Lagrange mean value 
theorem ensures a linear homogeneous relationship between these constitutive 
quantities and the related thermodynamic forces also in this mixed steady-state 
equilibrium case. 

The linearization of \re{ss_epr} result in the following 
expression:
\eqn{tDR1}{
\Delta \mu &= l_1 \ln\left(\f{V}{V_0}\right) - l_{12} \kappa \rv, \nl{tDR2}
\dot \rv &= l_{21} \ln\left(\f{V}{V_0}\right) - l_2 \kappa \rv.
}

We will call \re{tDR1}-\re{tDR2} as {\em thermodynamical aging law}. \re{tDR1} 
is identical to the Nakatani form of the constitutive law \re{Nak_claw} and 
\re{tDR2} is similar to \re{Nak_Relaw}.

The coefficient matrix may depend on the thermodynamic forces, in particular 
$l_1$, $l_2$, $l_{12}$ and $l_{21}$ may depend on the state variables $V$ and 
$r$  \cite{Gya77a}.  In the following we assume  a strict linear  
relationship, when the coefficient matrix is constant. It is remarkable, that 
there are no reasons to assume 
symmetry or antisymmetry of the matrix. The conditions of Onsagerian 
statistical background cannot be introduced without a particular 
interpretation of the internal variable (more detailed arguments are given in 
\cite{VanAta08a,VanEta14a}.

\section{Different mechanisms of different relaxations}

According to the experimental observations in case of velocity step and healing 
experiments, sometimes the internal variable changes when the surfaces slip, 
and sometimes its evolution is seemingly independent of the relative motion of 
the surfaces. This distinction is connected to the detailed mechanism of 
friction, and requires an extension 
of the modeling framework. In the following we will investigate the question of 
slip related internal variable evolution.

Ruina \cite{Rui83a} applied this assumption directly to the relaxation. In our 
case that requires the modification of \re{tDR2}, assuming that the slip is 
what makes change of the internal variable and the rate of the variable is a 
consequence. In this case the time derivative in \re{tDR2} is substituted by 
the space derivative and the rate is obtained as a consequence:  
\eqn{Ruicond}{
\f{d\rv}{dt} \rightarrow \f{d\rv}{d x} =  \frac{\dot \rv}{V}.
}
Performing this substitution in \re{tDR2} leads to
\eqn{tR2}{
\dot \rv = l_{21} V \ln\left(\f{V}{V_0}\right) - V l_2 \kappa \rv.
}
This equation, together with \re{tDR1} will be  called {\em thermodynamical 
slip law}.
One can see, that the Nakatani transformed  Ruina-law \re{Nak_claw}, 
\re{Nak_Relaw} can be obtained if $l_1=a$, $l_{12} = -1/\kappa$, $l_{21} =-b/L$ 
and $l_2 = 1/(\kappa L)$. The number of phenomenological parameters is 
increased by one, from three to four, compared to the original Ruina theory. It 
is because $\kappa$ appears only as a multiplier of the cross coefficients 
$l_{12}$ and $l_{21}$. The slip governed modification represents a particular 
quasilinear form of the strictly linear relations of the thermodynamic aging 
law \re{tDR1}-\re{tDR2}.\footnote{In our thermodynamical framework a slip 
related change may lead to further consequences. In order to keep the 
integrity of the thermodynamic considerations, the calculation of the entropy 
rate may be substituted by the calculation of the slip related entropy change. 

For example, with the internal energy, \re{Eintm}, the  
derivative of the entropy by the displacement will be the following:
\eqn{Sbalms}{
\frac{dS}{d x} =  \frac{1}{T} \left(F_e \frac{V_0}{V} + F_d  \frac{V-V_0}{V}
-k r \frac{dr}{dx} \right).
}
Then we proceed assuming logarithmic increment and obtain \re{tDR1} and 
\re{tR2} as a consequence. However, slip and displacement are not the same, the 
calculation of slip related changes should distinguish between the permanent 
and recoverable parts. Here we do not analyse this possibility further, we 
accept the approach of Ruina at this point.}


The slip condition of Ruina, expressed by \re{Ruicond} assumes that $V$ is the 
relative surface velocity, related to the permanent part of the displacement. 
When introducing a distinction between a permanent and a recoverable parts of 
the apparent displacement beyond the difference of the load point an relative 
surface velocities, one may expect, that only the permanent part contributes to 
the entropy production, to the evolution of the internal variable. For example 
an interpretation of the internal variable as recoverable displacement with the 
condition \re{plastkin} leads to the following permanent displacement:
$$
x_{per} = \int_{t_0}^t z(s)ds = x - r
$$

Particular mechanisms require that the internal variable influences the 
slip. For example when the internal variable is connected to deformation of 
surface irregularities, then this conclusion is straightforward. 
Therefore we assume in general, that the internal variable 
directly influences the displacement and the reduced part is what influences 
the (incremental) entropy production. Hence we introduce $x_{red} = x- 
\alpha \rv$, where $0\leq \alpha \leq 1$ is the {\em factor of slip reduction}. 
If $\alpha=1$, then the internal variable can be interpreted as 
the recoverable part of the displacement \cite{MitVan14a}. 

Therefore the evolution equation of the internal variable is obtained by 
substituting the time derivative with the slip related change of the internal 
variable, as follows:
$$
\f{d\rv}{dt} \rightarrow \f{d\rv}{d (x-\alpha \rv)} =  \frac{\dot 
\rv}{V-\alpha \dot\rv}.
$$
For the sake of simplicity $\alpha$ is constant. Then a simple rearrangement 
leads to the following evolution equation of the internal variable:
\eqn{ts1}{
\dot \rv &= V \f{l_{21} \ln\left(\f{V}{V_0}\right) - 
	l_2 \kappa \rv}{1+
	\alpha\left(l_{21} \ln(\f{V}{V_0}) - l_2 \kappa \rv \right)}.
}

Together with \re{tDR1}  we will call this  equation {\em thermodynamical 
friction law}. This is an 
other particular quasilinear form of the thermodynamic aging law 
\re{tDR1}-\re{tDR2}.

In the above thermodynamical models there is a direct effect and the 
conventional step test parameters are $a=l_1$ and $b=\f{l_{12}l_{21}}{l_2}$. 
In the following we compare the performance of the obtained thermodynamic models
with the classical models and experiments.

\section{Velocity step tests}

\begin{figure}[ht]
\centering
\includegraphics[width=11cm,height=8cm]{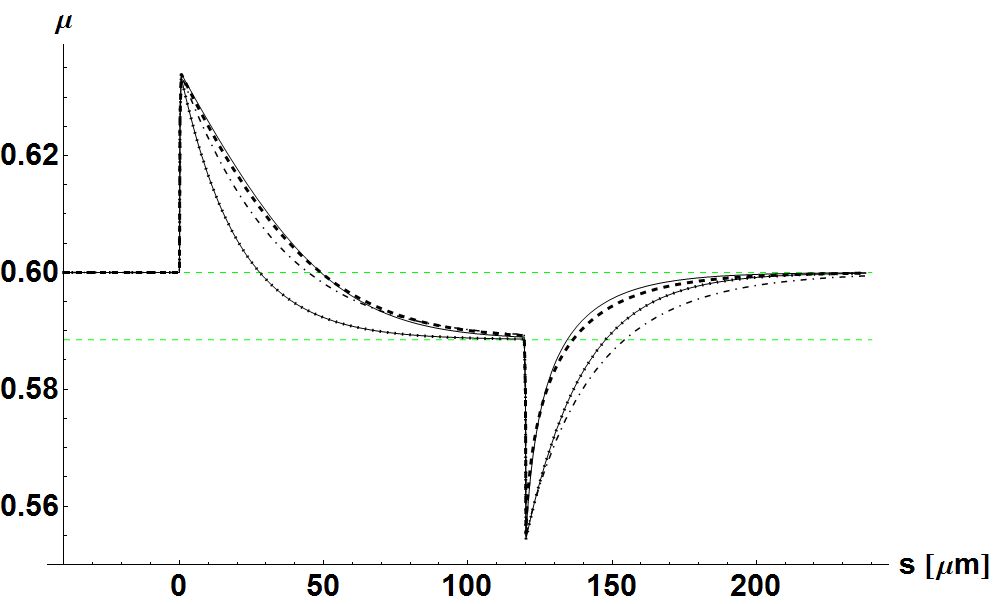}
\caption{Simulation of a velocity step test with different friction laws. 
The Dieterich and Ruina laws are drawn by solid thin lines. The thermodynamic 
friction 
law leads to the dotted line with $\kappa=1$ and $\alpha=0$ parameters running 
over the line of  Ruinalaw. The dashed line with parameters $\kappa=0.85$ and 
$\alpha=8$ runs 
close to the Dieterich law. For the dotdashed line the parameters are 
$\kappa=0.7$ and $\alpha=2$.}
\label{velojump_uni}
\end{figure}

A comparison of the different rate- and state dependent friction laws is 
shown in Figure \ref{velojump_uni}. The velocity weakening experiment is 
modeled with the following parameters:

$\mu_0=0.6$, $V_0 = 1\mu m/s$, $V_1 = 10\mu m/s$, 
 $L=20\mu m$, $a=0.015$, and $b=0.02$. 

Dieterich and Ruina models shown by a solid thin lines. The  parameters of the 
thermodynamic friction model are calibrated to give the proper step conditions 
and relaxation speed:  $l_1 = a$, $l_2 = 1/L$, $l_{12} = b$,  $l_{21} = l_2$. 
There are two additional parameters, the static recovery strength $\kappa$, 
analogous to a spring constant, and the factor of slip reduction $\alpha$. 
The dotted curve runs exactly over the line obtained by the Ruina model, 
because thermodynamic slip model recovers the Ruina case when $\kappa=1$ and 
$\alpha=0$. One obtains highly asymmetric relaxation curves choosing higher 
values of the slip reduction parameter $\alpha$ and with a proper choice one 
calculates curves that run close are pretty close to the Dieterich model. E.g. 
the thick dashed curve was calculated with $\kappa=0.85$ and $\alpha=8$. 
Moreover, one can obtain 
symmetric relaxation curves close to either the down or up relaxation curves of 
the Dieterich model with an appropriate choice of the parameters. For example 
the thick dotdashed line was calculated using $\kappa=0.7$ and $\alpha=2$. 

The thermodynamic aging law produces asymmetric curves, similar to the 
Dieterich law.

\section{Healing}

\begin{figure}[ht]
\centering
\includegraphics[width=11cm,height=8cm]{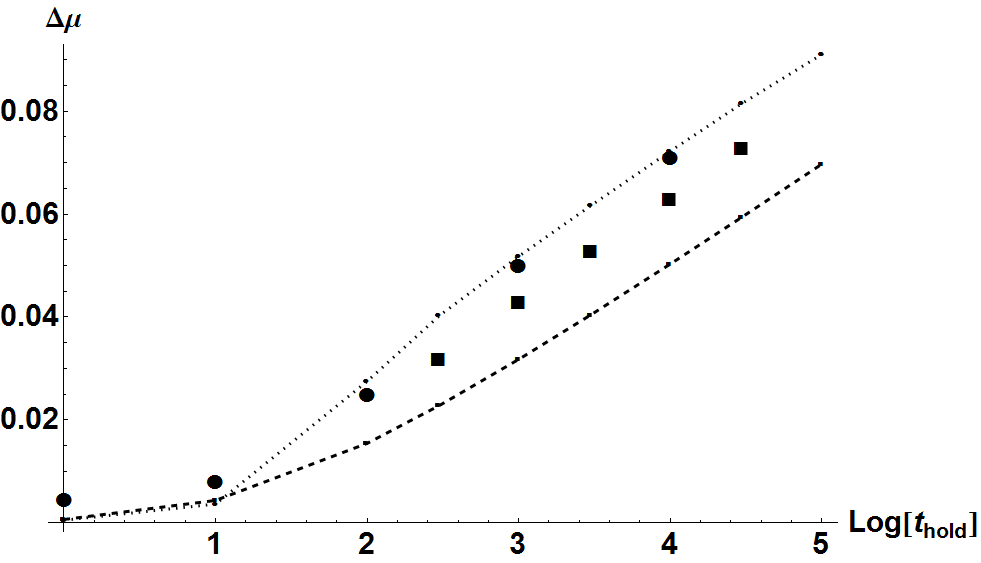}
\caption{Simulation of the healing experiments of Beeler et al 
\cite{BeeEta94a}. The experimental data for the $k=0.002$ case is shown by the 
big circles and for 
the $k=0.074$ case by the rectangles. The dotted curve is calculated by the 
thermodynamic friction model for $k=0.002$. The  $k=0.074$ parameter 
resulted in the dashed curve.}
\label{heal0}
\end{figure}

The interpretation of healing is contradictory, therefore we have chosen the 
experiments and strategy of Beeler an Tullis \cite{BeeEta94a,Maro98a} for 
demonstration. They performed healing experiments with different machine 
rigidity. One of them $k=0.002$ was the natural elasticity of the experimental 
device, in this case the load point velocity was $V_0=1 \mu m/s$. With the  
$k=0.074$ servocontrolled value the load point velocity was $V_0=0.316 
\mu m/s$. In both cases the rock parameters were $L=3\mu m$, $a=0.009$, and 
$b=0.008$. The data points in the first and second cases are shown on figure 
\ref{heal0} by circle and rectangles, respectively. The simulation used the
experimental rock parameters of Beeler et al. \cite{BeeEta94a}  and the 
additional parameters 
were chosen as $\alpha = 0.75$ and $\kappa = 0.9$. Then we have obtained the  
dashed curve for the $k=0.074$ case and the dotted one for $k=0.002$. 

\begin{figure}[ht]
\centering
\includegraphics[width=11cm,height=8cm]{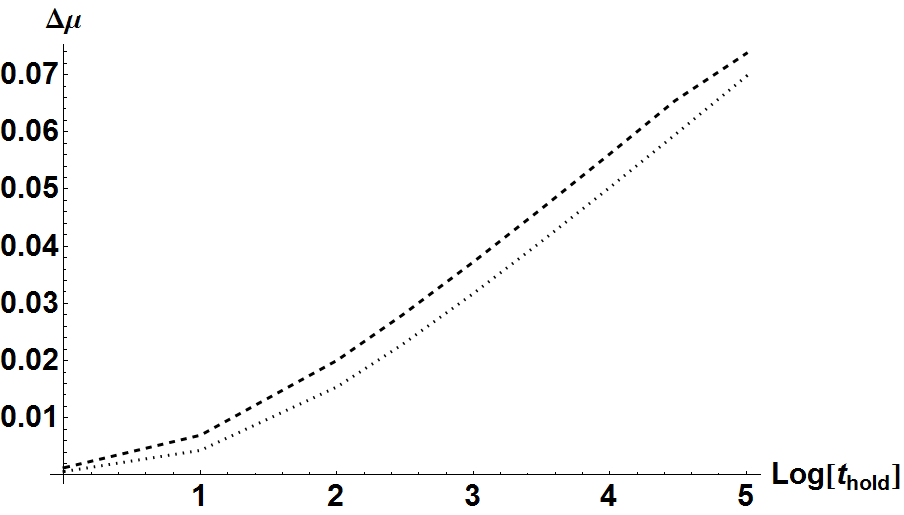}
\caption{The effect of larger steade state velocity for healing. $V_0= 1\mu 
m/s$, dotted curve, $V_0=2\mu m/s$, dashed one. }
\label{heal1}
\end{figure}

\begin{figure}[ht]
\centering
\includegraphics[width=11cm,height=8cm]{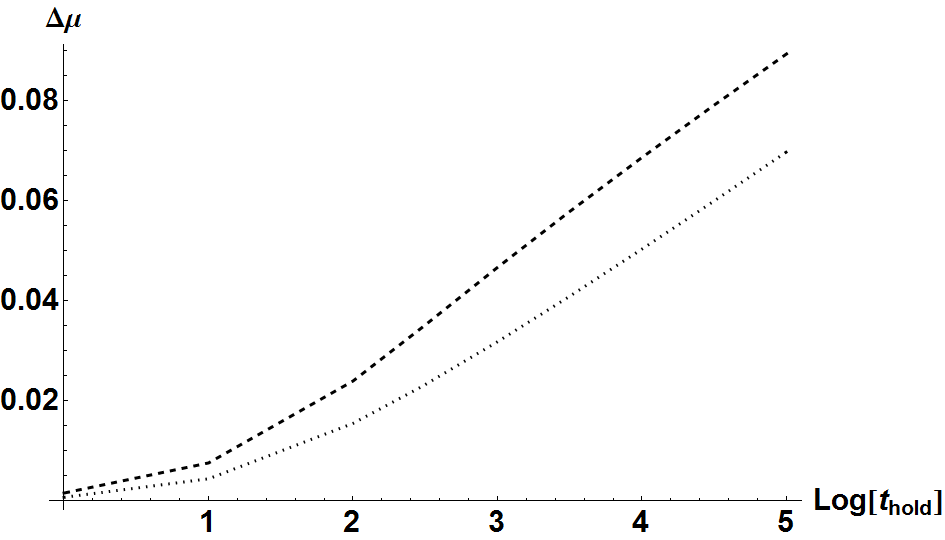}
\caption{ The effect of softened machine for healing. $k= 0.002$, dotted curve, 
$k= 0.02$, dashed one.}
\label{heal2}
\end{figure}

Figures \ref{heal1} and \ref{heal2} show the effect of changing the initial 
velocity and the machine rigidity. In figure \ref{heal1} the simulation of the 
healing experiment with the more rigid machine is shown, where the parameter 
values are $k=0.002$,  $V_0=1 \mu m/s$, $L=3\mu m$, $a=0.009$, $b=0.008$,  
$\alpha = 0.75$ and $\kappa = 0.9$ (dotted curve in figure \ref{heal0}). 
Increasing the velocity to  $V_0=2 \mu m/s$ pushes the curve upward and 
parallel to the original one, shown with the dashed curve. In figure 
\ref{heal2} the effect of softening is 
demonstrated, the dashed curve is calculated with $k=0.02$. The dotted curve  
is identical to the one in figure \ref{heal1}.  The increase of the 
healing effect qualitatively corresponds to the experimental observations. 


\section{Summary and conclusion}

A non-equilibrium thermodynamic model with a single internal variable of rate- 
and state-dependent friction was proposed. The model introduced the following 
basic assumptions:
\begin{itemize}
\item The deviation from the steady state is logarithmic.
\item The change of internal variable is due to a reduced slip, and the 
reduction is proportional to the value of the internal variable.
\end{itemize}
The obtained thermodynamic friction model generalizes the well known 
Dieterich and Ruina laws with two additional parameters.
\begin{itemize}
\item For velocity step tests it interpolates between the Dieterich
and Ruina laws. The form of the relaxation depends on the additional 
parameters. One can obtain symmetric up and down relaxation curves that 
are either close to the curves of Ruina relaxation or
are close to the Dieterich type relaxation, but with less apparent linear part.
\item Simulations show promising results for healing experiments, both 
quantitavely and qualitatively.
\end{itemize}

\section{Acknowledgement}   
The work was supported by the grants Otka K81161 and K104260. 


\end{document}